# Holographic Colour Prints: Enhanced Optical Security by Combined Phase and Amplitude Control

Kevin T. P. Lim[1,†], Hailong Liu[1], Yejing Liu[1], Joel K. W. Yang[1,2,*]

[1] Singapore University of Technology and Design (SUTD), 8 Somapah Road, Singapore 487372

[2] Institute of Materials Research and Engineering (IMRE), 2 Fusionopolis Way, Innovis, #08-03, Singapore 138634

[†]*Present affiliation: Cavendish Laboratory, JJ Thomson Ave, Cambridge CB3 0HE, United Kingdom*

*Corresponding author: joel_yang@sutd.edu.sg*

## Abstract

We created a novel optical security device that integrates multiple computer-generated holograms within a single colour image. Under white light, this "holographic colour print" appears as a colour image, whereas illumination with a red, green, or blue beam from a handheld laser pointer projects up to three different holograms onto a distant screen. In our design, all-dielectric layered pixels comprising phase plates (phase control) and structural colour filters (amplitude control) are tiled to form a monolithic print, wherein pixel-level control over the phase and amplitude of light allows us to simultaneously achieve hologram multiplexing and colour image formation. The entire print is fabricated in a single lithographic process using a femtosecond 3D direct laser writer. As the phase and amplitude information is encoded in the surface relief of the structures, our prints can be manufactured using nanoimprint lithography and could find applications in document security.

Optical security devices are valuable tools in data encryption and document authentication as they exploit the many properties of light, including amplitude, phase, polarisation, and wavelength, to achieve distinctive visual effects that can be difficult to decode or duplicate.[1,2] The two archetypal optical security devices are microscopic colour prints[3–5] and holograms[6]. Microscopic colour images can be directly viewed under a magnifying glass, whereas holograms are easily verified by using a laser pointer to project an image onto a screen placed in the far field (Fraunhofer regime). To improve the security of these basic devices, additional complexity is usually introduced by encoding multiple sets of information into a single device, i.e. multiplexing.

Multiplexed colour prints have been created by encoding information in two independent dimensions of elongated metal nanostructures, allowing for two different images to be read out under orthogonal polarisations of light.[7,8] Using similar nanostructures of various sizes optimised to respond to different wavelengths, three-colour multiplexed holograms based on the Pancharatnam-Berry (PB) geometric phase have been demonstrated.[9–11] Multiplexed PB holograms have also been fabricated using an alternative geometry of nanoslits in a metal film.[12,13] Unfortunately, transmission PB holograms often suffer from low transmission efficiency due to their use of lossy metal nanostructures and are also complicated to read out, requiring the use of circularly polarised light as well as specific illumination and/or viewing angles. Additionally, the nanostructures are fabricated with electron beam lithography or focused ion beam milling, which incurs high costs and imposes practical constraints on the patternable area. These shortcomings, namely low transmission efficiency, complexity of readout, and high fabrication costs, have limited their practical application in optical security devices thus far.

In comparison, traditional phase elements consisting of dielectric structures of different thicknesses ("phase plates") enable holographic projection to be achieved with higher transmission efficiency, simpler illumination methods (e.g. a handheld laser pointer), and little restriction on the polarisation or incidence angle of the light. They are also potentially easier and cheaper to manufacture than PB nanostructures as their larger dimensions are within the resolution limit of photolithography. High transmission efficiency multiplexed holograms that project up to three different images depending on the incident wavelength have previously been demonstrated using a variety of techniques including phase modulation[14–16] and depth division[17]. Recently, white-light transmission colour holograms[18] operating in the Fresnel limit have also been developed.

However, as phase holograms are not designed to control the amplitude of light, they generally appear random or featureless under incoherent illumination, which makes them less attractive as optical security devices. Conversely, colour images have superior decorative values on banknotes or passports but generally cannot produce any meaningful holographic projection under coherent illumination as they do not control the phase of the light. Introducing a design methodology to control the phase and amplitude of light simultaneously is an area that has been relatively unexplored and could enable the creation of a dual-function device that appears as an image in plain view, but encrypts additional data that can be retrieved through holographic projection.

Here, we propose a new type of optical security device that combines phase and amplitude control to integrate (multiple) holograms into a colour print, which we refer to as a (multiplexed) holographic colour print. Our device appears as a colour image when viewed in white light, but reveals up to three different hidden grayscale holographic projections under red, green, and blue laser illumination. To the best of our knowledge, this is the first time that multiple holograms have been encrypted into a colour print. Doing so requires the ability to encode both phase and colour independently within individual pixel elements, which is a challenge for recent PB holograms. Our holographic colour prints provide a unique and easily recognisable visual effect, and may be of interest to the security industry as effective anti-counterfeiting elements that provide enhanced optical security on important documents such as banknotes and passports.

Figure 1 illustrates the concept of the (transmission) holographic colour print. The top layer contains colour filters that encode a colour print, and the bottom layer contains phase plates that encode the holograms. The colour filters have two functions: (1) to collectively form a colour image under white light illumination, and (2) to control the transmission of red, green, and blue (RGB) laser light through the pixels of the underlying multiplexed holograms. With coherent monochromatic illumination (e.g. light from a laser pointer), the incident light is filtered such that only the relevant phase plates with colour filters that closely match the illumination wavelength are selected for a given holographic projection, while the other phase plates do not form a projection as their colour filters are mismatched and reject the incident light. Therefore, the multiplexed holographic colour print will show different holographic projections when illuminated by red, green and blue lasers. Because incident light passes through all pixels in parallel, the pixels can act independently to allow transmission of certain wavelengths in some regions of space but not others, which enables several holograms to jointly occupy the total area available in a spatial multiplexing scheme. By exploiting the freedom to divide the space into regions of arbitrary shapes and sizes, the individual hologram areas can then be strategically allocated such that the arrangement of their colour filters additionally encodes a chosen colour image. Under incoherent white light illumination, the phase modulation of the holograms is effectively ignored and the colour filters act as amplitude-modulating colour pixels that together show the desired colour image.

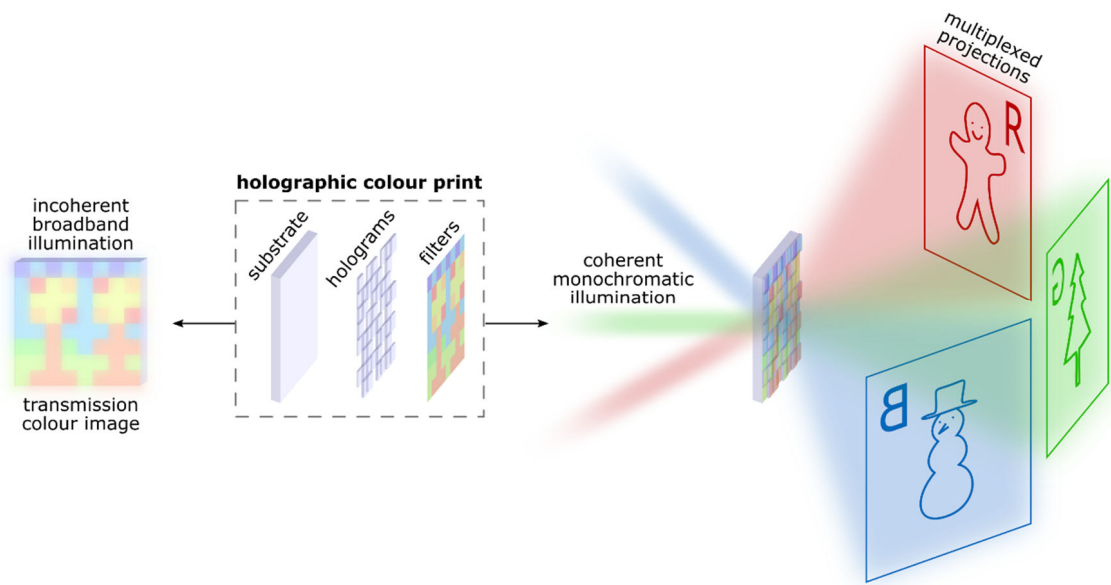

***Figure 1. Concept and working principle of a holographic colour print.*** *Exploded-view schematic of a transmission holographic colour print (centre), a layered optical device in which colour filters are integrated on top of holograms. The colour filters act as colour pixels in a colour image under white light illumination (left) and also serve to control the transmission of red, green, and blue (RGB) light through the pixels of the underlying multiplexed holograms (right). Under RGB laser illumination, each wavelength of light selects a different holographic projection, which is independent of the colour image and the other projections. Far field projections appear on-axis at normal incidence and off-axis at angled incidence. Three different angles of incidence are shown to illustrate the three distinct holographic projections. The projections remain in focus at any distance in the far field, can be achieved over a wide range of incident angles, and are perfectly overlapped under collinear multi-colour illumination.*

**Results**

Design of the holographic colour pixel

To create a physical realisation of a holographic colour print, we first developed a holographic colour pixel that controls both the phase and amplitude of light. As our pixel has a relatively large minimum feature size of several hundreds of nanometres and consists entirely of a single dielectric material, we are able to fabricate it with femtosecond 3D printing (direct laser writing) as a monolithic structure in a cross-linked polymer (see Methods section for details of the fabrication process and materials used).

Our pixel design (Fig. 2a) integrates a dielectric phase plate under a structural colour element comprising an array of dielectric pillars, which acts as a colour pixel for the transmission colour image under white light illumination and also as a colour filter[19] to selectively transmit red, green, or blue laser light. The dielectric phase plate controls the phase of transmitted light according to the equation $\phi(\lambda) = 2\pi(n-1)t/\lambda$, where the phase shift ($\phi$) arises from path length differences that depend on the phase plate thickness (t) and refractive index (n). The refractive index of the dielectric polymer material we used (between 1.54 and 1.58 in the visible region)[20] allows us to achieve a full $2\pi$ phase modulation for red, green, and blue light by varying the phase plate thickness over a range of 1.2 μm. Within this thickness range, we define 11, 9, and 7 discrete thickness levels for $2\pi$ phase modulation to realise red, green, and blue holograms respectively (details in Supplementary Information).

Assuming that a phase plate acts as an ideal phase-controlling (constant-amplitude) element and a pillar array colour filter acts as an ideal amplitude-controlling (constant-phase) element, combining these elements into a layered pixel should allow for independent phase and amplitude control. In practice, however, due to the refractive index mismatch between the glass substrate and the polymer structures, changing the thickness of the underlying block to control the phase also affected the transmission amplitude of the overall pixel. The shift in transmission spectrum with block thickness caused a significant change in the pixel colour with thin blocks of t = 0–0.4 μm, but was not noticeable for thicker blocks of t ≥ 0.6 μm (see Supplementary Fig. 1). As such, we chose to work with blocks of 0.6–1.8 μm thickness to span the required range of 1.2 μm.

To compensate for the remaining variations in pixel transmission amplitude due to differences in phase plate height, we fabricated dielectric pillar arrays on blocks with thicknesses varying between 0.6 to 1.8 μm and measured their transmission spectra $T(\lambda)$. We could then average out the dependence of $T(\lambda)$ on thickness, effectively eliminating any residual phase-amplitude coupling. Subsequently, varying the pillar array dimensions of height (*h*), diameter (*d*), and pitch (*p*) gave us access to a range of colours (see Supplementary Fig. 1) from which we selected the most suitable filters for red, green, and blue wavelengths. The transmission spectra $T(\lambda)$ of the optimised RGB colour filters (Fig. 2b) show a reasonably good wavelength selectivity (i.e. mutually exclusive, or orthogonal, transmission at the wavelengths of interest), as all the filters have a high transmittance of ~60% at the desired wavelength and a low transmittance of ~20% at unwanted wavelengths. The wide colour range enables us to select suitable colours to reproduce the colour image under white light illumination, while the good wavelength selectivity ensures that laser light can be filtered to distinguish the individual holograms.

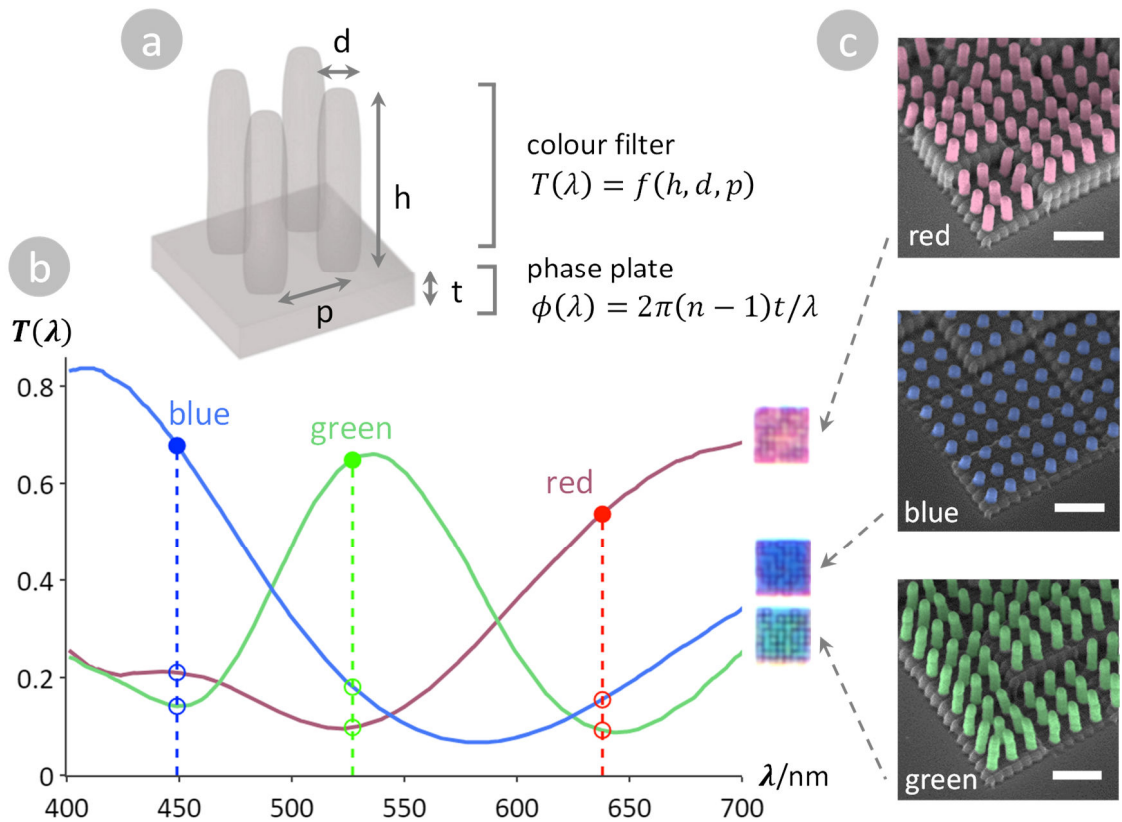

*Figure 2. Structure and characteristics of an all-dielectric holographic colour pixel. (a) Schematic of a holographic colour pixel that provides combined phase and amplitude control, comprising an array of pillars (colour filter) integrated on top of a block (phase plate) in a dielectric with refractive index $n$. The colour filter controls the amplitude of light based on its transmission spectrum $T(\lambda) = f(h, d, p)$, which depends on the pillar array dimensions $\{h, d, p\}$ (height, diameter, and pitch). The phase plate controls the phase of transmitted light where the phase shift arises from path length differences that depend on the block thickness. (b) Transmission spectra and corresponding optical micrographs of pillar arrays with red, green and blue colours. Transmission spectra were averaged from measurements of pillar arrays with the same dimensions as those shown in the images, but patterned separately on blocks of uniform thickness (0.6, 1.0, 1.4, 1.8 μm). Good RGB wavelength selectivity can be seen from the high transmittance of ~60% for red (638 nm), green (527 nm), and blue (449 nm) light passing through their respective filters (filled circles), and the low transmittance of ~20% for light passing through the wrong filters (empty circles). (c) False-colour tilt-view SEMs of pillar arrays with dimensions optimised to give the best selectivity for red, green, and blue (RGB) light.*

*The pillars (~400 nm in diameter and respectively 1.9, 0.7, and 2.6 μm in height) are patterned in a square array of 1 μm pitch onto 3×3 μm² blocks of randomly varying thickness within the thickness range to be used for hologram phase plates (0.6–1.8 μm). Scale bars are 2 μm.*

Fabrication of holographic colour prints

With our holographic colour pixel design, it is possible to create multiplexed holograms by fabricating large arrays of pixels. In the simplest case, holograms can be multiplexed side-by-side with the phase plates of each hologram spatially segregated in contiguous singly-coloured regions, giving a similar result to that achievable by pasting macroscopic colour filters onto a spatial light modulator[21]. However, this multiplexing scheme is incompatible with realising an arbitrary colour image. In the design of our holographic colour print, the ability to control phase and amplitude on the level of individual pixels (i.e. pixel-level control) grants us the freedom to move pixels around as long as the phase is recalculated for any new pixel arrangement (Fig. 3a). As the total area assigned to each hologram is not fixed, pixels of one hologram can be replaced with pixels of another, which means that we are free to choose near-arbitrary pixel arrangements (see Supplementary Information for the limits of this freedom). Crucially, replacing pixels, rearranging amplitudes, and recalculating the phases (see Methods section for details of our design algorithm) does not greatly affect the fidelity of holographic projection when the number of pixels is large (e.g. 480×480 pixels, as used here).

Exploiting this spatial degree of freedom in a two-tone multiplexed hologram allows us to arrange the pixels to form a meaningful binary image such as a QR code without a significant decrease in the fidelity of the holographic projections, as shown in Fig. 3b. Although these holograms are designed for red and blue laser illumination, it is worth mentioning that the colour filters do not necessarily need to be red and blue as long as their transmission amplitudes are mutually exclusive (orthogonal) at the laser wavelengths used. To improve the visibility of the QR code for scanning under broadband white light illumination, the colours yellow (high average transmittance of 48% in the range 450–650 nm, which contains most of the power of a typical white light source) and blue (low average transmittance of 22%) are chosen for the light and dark QR code pixels respectively to maximise the grayscale image contrast, while still retaining wavelength selectivity for the multiplexed holograms under monochromatic red and blue light. This can be seen as an additional degree of freedom as we are able to choose colour filters with transmission spectra that best match both the desired image colours and the hologram design wavelengths.

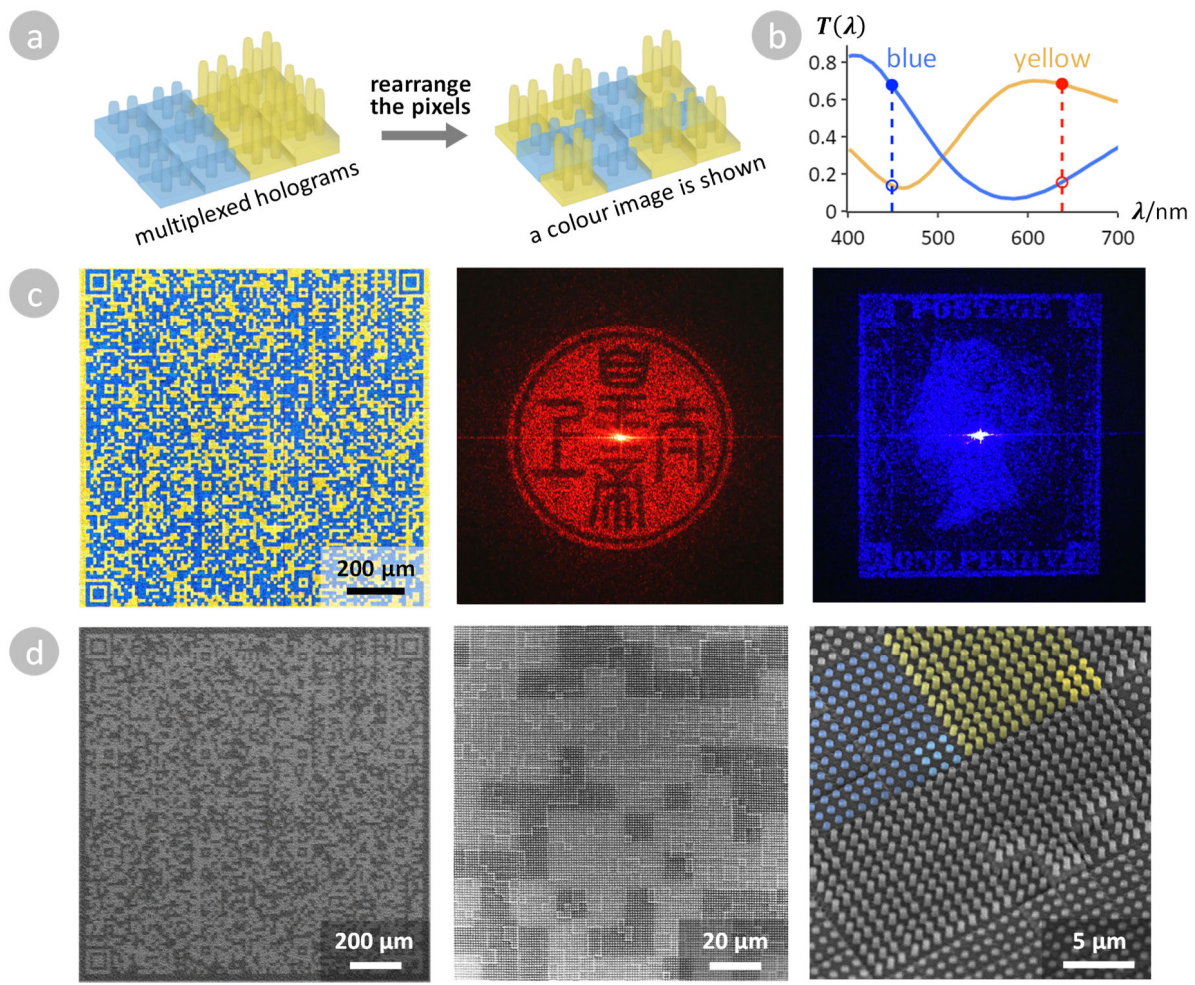

***Figure 3. Demonstration of a two-tone holographic colour print.** **(a)** Design principle of the holographic colour print: The ability to choose the amplitudes of the pixels independent of their phase makes it possible to rearrange the pixels in a multiplexed hologram ("spatial freedom") so that the pixel arrangement contains meaningful information as well, allowing a colour image to be shown on top of the holograms. As long as the phase is calculated taking into account the pixel arrangement, the individual holographic projections can still be maintained.* ***(b)*** *Transmission spectra of the yellow and blue colour filters used, showing mutually exclusive (orthogonal) transmission at the wavelengths of interest (638 nm red and 449 nm blue).* ***(c)*** *Optical characterisation of the print: (left) Transmission optical micrograph of a two-tone multiplexed hologram in which the pixels are arranged to form a QR code colour image. The blue colour filters (blue hologram channel) selectively pass blue light but not red light, whereas the yellow colour filters (red hologram channel) selectively pass red light but not blue light. Holographic projections in transmission, photographed on a white wall in a dark room:*

*(centre) the image of a Chinese seal is shown under 638 nm red laser illumination and (right) the image of a Penny Black stamp is shown under 449 nm blue laser illumination. The projections are approximately 10 cm large when projected at a distance of 1 m.* **(d)** *Scanning electron micrographs (SEMs) of the print at various scales. Each holographic colour pixel consists of a 3×3 pillar array on top of a 3×3 µm$^2$ block, and each QR code pixel is a super-pixel comprising a 4×4 block of holographic colour pixels. In the close-up tilt-view SEM, a blue and a yellow QR code super-pixel are highlighted in false colour, and the bottom-right corner holographic colour pixel of each is further highlighted.*

Using this spectral degree of freedom, we relaxed the constraint on wavelength selectivity of the colour filters by introducing three additional colours (orange, yellow, and purple) into a three-colour RGB (red, green, and blue) hologram and arranged the pixels to form a complex six-colour image (Fig. 4a). The additional colours were assigned to their closest match within RGB: orange colour filters were placed over phase plates belonging to the red hologram, and purple colour filters over phase plates of the blue hologram. As the selectivity of yellow colour filters for transmitting red light over green light was too poor, no hologram information was stored in phase plates under yellow filters – it would otherwise cause the red projection to appear in the green projection as "crosstalk". Instead, a random phase was stored in the yellow pixels so as to diffuse their contribution to the transmitted zero-order (undiffracted) beam, minimising contamination of the other projections with crosstalk from the yellow pixels. In the final six-colour print, the high fidelity of holographic projections and remarkable lack of discernible crosstalk (Fig. 4c) demonstrate that it is possible to pattern complex and colourful images without sacrificing the quality of the multiplexed holograms in the same print. We note that the choice of image allowed for the use of dithering in recolouring the image to obtain an optimal pseudo-random pixel arrangement (details in Supplementary Information), affording holographic projections superior to those from the QR code hologram in Fig. 3c.

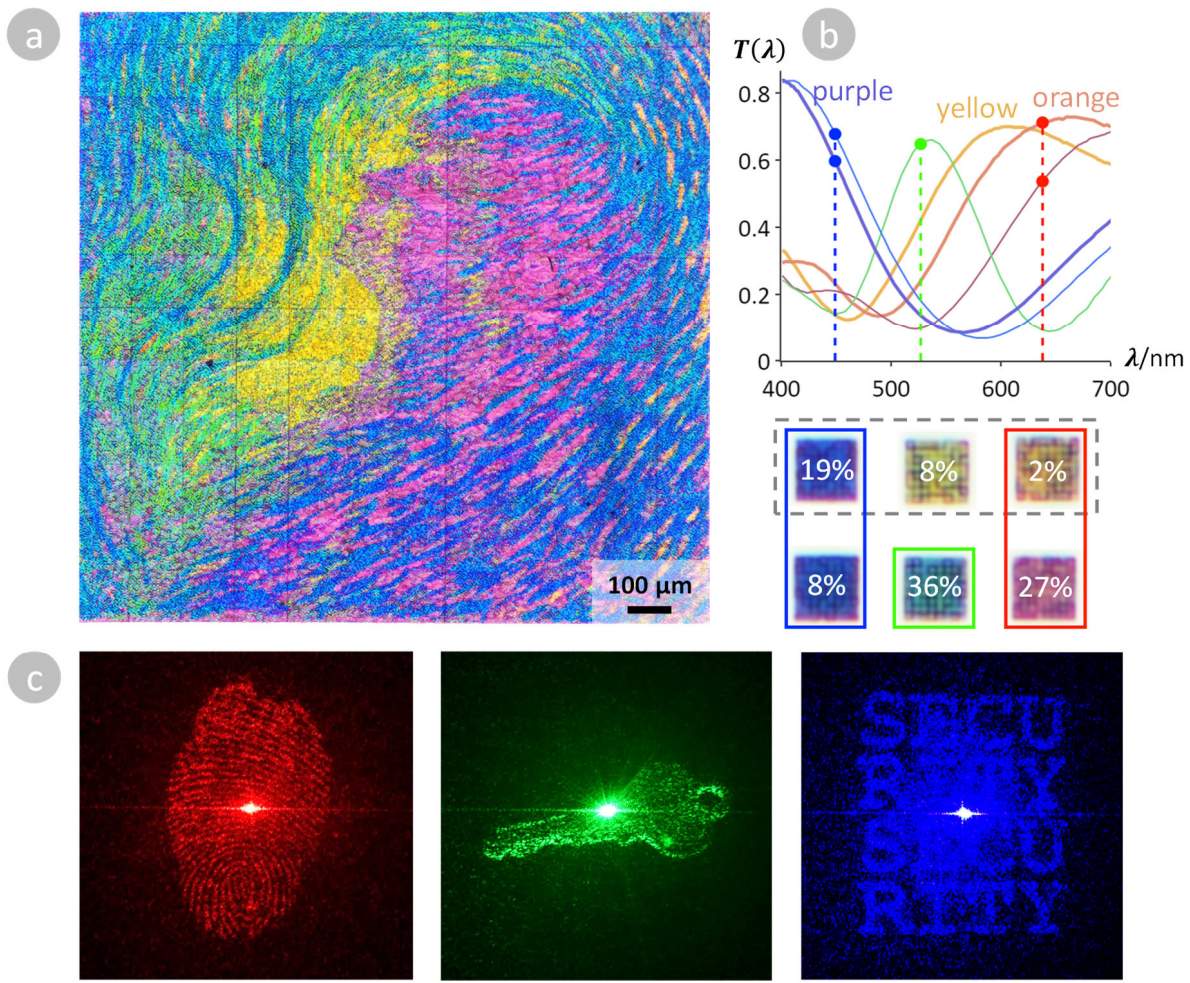

*Figure 4. Enhanced optical security provided by a six-colour holographic colour print. (a) Transmission optical micrograph of the colour print, a reproduction of Luigi Russolo's painting* Perfume *under which three holograms have been multiplexed. The ability to include colours that are less suitable for multiplexing the holograms ("spectral freedom") allowed us to expand the usable colour palette to a total of six colours. (b) Transmission spectra and corresponding optical micrographs of pillar arrays that produce the colours used in the print: added colours orange, yellow, purple (spectra plotted with thick lines; micrographs marked with a dashed box) and original colours red, green, and blue (spectra plotted with thin lines). Overlaid percentages on the micrographs of each pillar array denote the proportion of pixels in the print with that colour. Coloured boxes around the pillar micrographs indicate that they are used in the hologram channel of that colour. (c) Holographic projections of the print in transmission, photographed on a white wall in a dark room: (left to right) a red thumbprint, a green key and blue lettering that reads "SECURITY". Illumination sources were*

*638 nm red, 527 nm green, and 449 nm blue lasers respectively. The projections are approximately 10 cm large when projected at a distance of 1 m.*

**Discussion**

Although dithering was used in the *Perfume* print (Fig. 4) to improve the holographic projections without significantly affecting the quality of the colour image, this could not be done for the QR code print (Fig. 3) as dithering locally scrambles the colours and positions of the pixels across the entire image, which would make it impossible to scan the QR code. Instead, we faithfully reproduced the QR code's super-pixel structure, with the resulting trade-off being that its holographic projections were blurred by convolution with the Fourier transform of the semi-periodic QR code super-pixel structure (details in Supplementary Information).

The choice of images ("Chinese seal" and "Penny Black") for the QR code holograms also compounds the problem of blurriness: (1) images with predominantly dark-on-bright contrast (Chinese seal) tend to be less tolerant to bright speckle noise in holographic projections which can easily obscure edges and other important features, (2) images with both foreground and background components (Penny Black) are complex, making it difficult to distinguish the subject from its surroundings in a noisy reproduction, and (3) projections containing a large total area of bright regions (both images) are inherently noisier as control of additional points of light can only be achieved at the expense of noise suppression given a fixed number of degrees of freedom. In light of these problems, we emphasise that the QR code holographic projections shown in Fig. 3 should be understood as a worst case scenario for the performance of holographic colour prints.

With our new pixel design that combines phase plates and structural colour filters, combined phase and amplitude control can be achieved to simultaneously fulfil the objectives of colour image formation and hologram multiplexing. Together with an algorithm that accounts for the transmittances of a range of colour filters, this enabled the creation of monolithically integrated holographic colour prints that operate as passive standalone devices capable of showing a colour image and multiple holographic projections. We anticipate that this method of exploiting phase and amplitude control at the level of individual pixels could lead to novel and practical optical security devices.

**Methods**

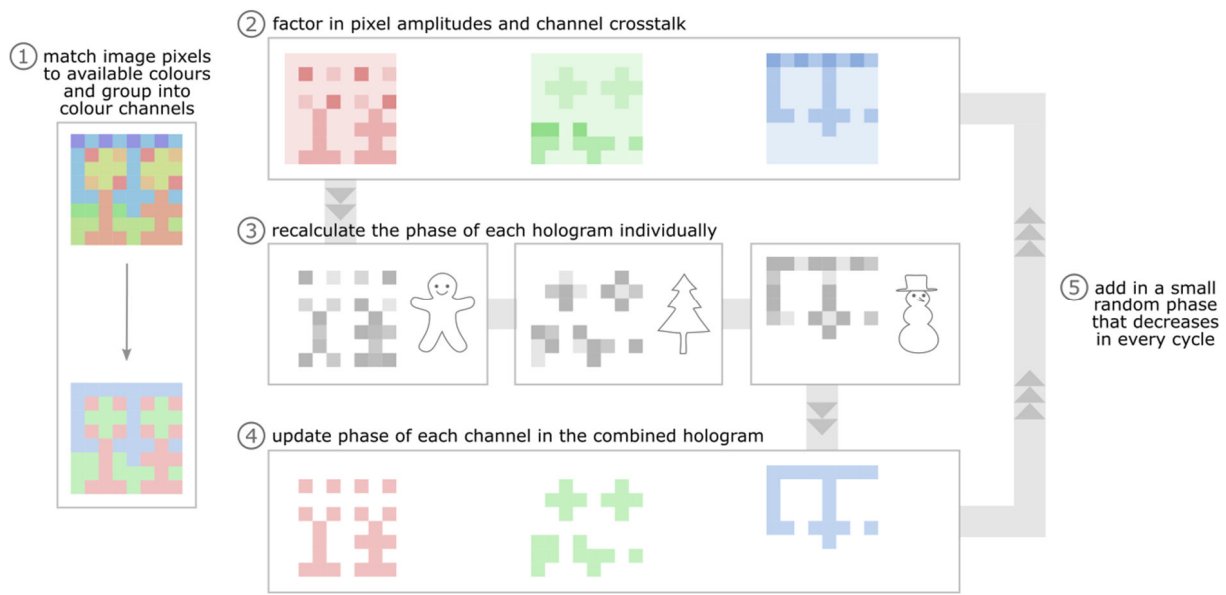

*Figure 5. Designing a holographic colour print. Flowchart of the design algorithm for combining colour image formation and spatial multiplexing of holograms. The initial stage of the algorithm (Step 1) recolours the input colour image using a limited colour palette and then divides the colour pixels into several groups (channels) based on their suitability for filtering each hologram. After the assignment in Step 1, the main body of the algorithm (Steps 2–5) applies a modified Gerchberg-Saxton algorithm that takes into account the arrangement of pixels as well as their amplitudes (transmission spectra) and phases in order to iteratively re-optimise the phase of the holograms on each channel. Despite the imperfect selectivity of the pixel amplitudes (there is non-zero transmission at unwanted wavelengths, which results in crosstalk of the signal between channels), a satisfactory balance between the quality of the colour image and holographic projections can be achieved by using a sufficiently large number of pixels, which allows for both spatial (as exemplified by Fig. 3) and spectral (as in Fig. 4) degrees of freedom to be exploited.*

An iterative multi-objective MATLAB code was written to perform the colour image matching and phase calculation for each hologram channel (Fig. 5). Step 1 takes as input data a set of microscope images and transmission spectra collected from many different colour filters consisting of pillar arrays with varying dimensions (the spectra used were averaged from colour filters on blocks of several thicknesses, as described in Fig. 2). This data provides the colour as well as transmittance (amplitude) of the pillar arrays at specific desired wavelengths. In Step 1, each pixel of the colour image to be patterned is colour-matched to the closest available colour in the dataset while balancing two considerations: the majority of the pixels in the image should have colours that are suitable for filtering RGB wavelengths, and the number of unique colours should be minimised to keep patterning time short. Once the colour filters are selected, the colour image is recoloured accordingly and the corresponding transmission spectra are used to generate a map of amplitudes (transmittance of each pixel at R, G, and B wavelengths). This information is then fed to a modified Gerchberg-Saxton[22] algorithm (Steps 2–5) to iteratively determine the phase of each element that can best achieve three separate R, G, and B grayscale holographic projections. We used a size of 480×480 pixels for our holographic colour prints, which could typically be designed in under a minute by running our code on a modern laptop.

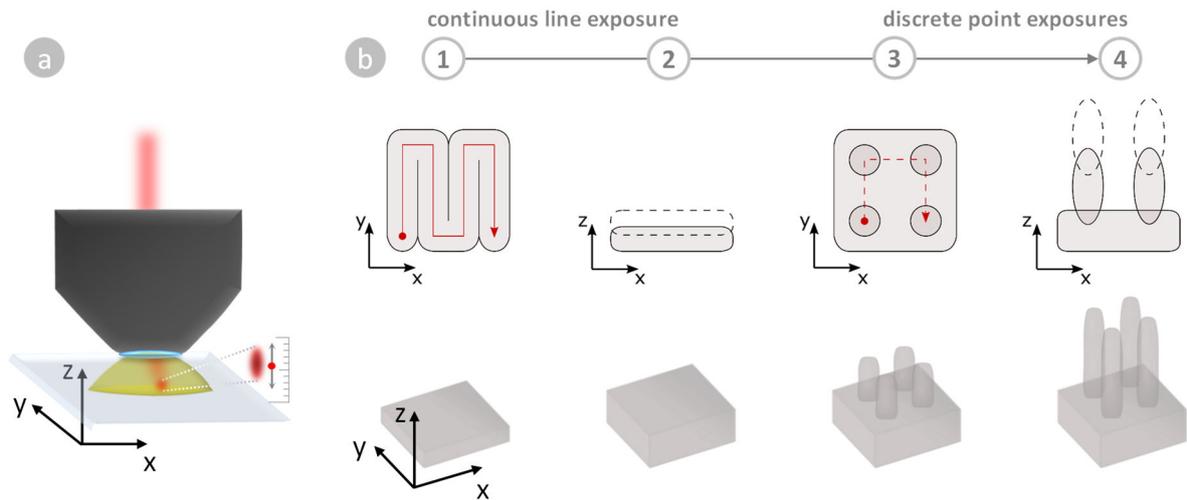

*Figure 6. Fabricating a holographic colour pixel. (a)* In the direct laser writing exposure process, an IR laser is focused into a liquid drop of negative-tone photoresist. At the focal point of the laser spot, the UV-sensitive photoresist is cross-linked by two-photon polymerisation and becomes solid. (Unexposed photoresist remains as a liquid and is later washed away during development.) The laser spot is translated in three dimensions (x, y, and z) according to the sequence in (b) to create complex structures. Resolution in the z-direction is not limited by the axially elongated shape of the point spread function as it is determined by the positioning accuracy of the laser spot. *(b)* Process flow for fabricating a holographic colour pixel using direct laser writing: (1) Blocks are created by rastering the laser spot to fill a square in the xy-plane with a continuous line exposure ("hatching"); (2) Step 1 is optionally repeated at higher z-positions ("slicing"); (3) Pillar arrays are created by point exposures, where the diameter is controlled by the exposure dose; (4) Step 3 is optionally repeated at higher z-positions. Block thickness and pillar height are controlled in Steps 2 and 4 by overlapping multiple layers of exposures along the z-direction. A full holographic colour print can contain many pixels with different block and pillar dimensions.

From the phase and amplitude maps created by the design algorithm in Fig. 5, a separate MATLAB code generated a blueprint of structures (phase plates and pillar array colour filters) with appropriate dimensions to achieve the desired phase and amplitude. This structural blueprint was finally converted into a set of instructions for controlling the laser writing sequence used in the fabrication process.

We fabricated holographic colour prints consisting of phase plates and colour filters in a single lithographic process by using 3D direct laser writing. A femtosecond pulsed IR laser is focused by a high numerical aperture immersion microscope objective into the photoresist as a tight spot of submicron size. Two-photon absorption and polymerisation occur in the UV-sensitive photoresist at the focal point of the laser spot, which can be scanned laterally and shifted axially (refocused) relative to the photoresist/glass interface according to a predefined writing sequence to write a desired pattern consisting of points and lines (Fig. 6). Cross-linking of the negative-tone photoresist along the laser exposure path creates the phase plates and colour filters as solid polymer structures on the glass.

The area to be patterned was split into a square grid of $120 \times 120$ $\mu m^2$ write-fields based on the maximum undistorted field of view of the objective lens. In each write-field, we grouped blocks of the same thickness and performed the writing sequence in Fig. 6b for each group in ascending order of thickness. Within each group of blocks, all blocks were patterned before their pillars were patterned. We chose to group the blocks by thickness instead of spatial coordinate to minimise patterning time, as refocusing in the z-direction is much slower than lateral scanning in the xy-plane. Further details of experimental parameters and equipment used are given in the Supplementary Information.

**Acknowledgements**

We acknowledge funding support from the SUTD Digital Manufacturing and Design (DManD) Center grant No. RGDM1530302, and the National Research Foundation grant award No. NRF-CRP001-021. We thank Junjie Kang for technical assistance in assembling a setup for collinear RGB illumination, Eric Tan for photography of the holograms, and Qiu Cheng-Wei for useful discussions.

**Author Contributions**

K.T.P.L. devised the concept, designed the code, performed experiments, and wrote the manuscript. H.L. and Y.L. performed experiments. J.K.W.Y. edited the manuscript and supervised the research.

**Data Availability**

The source images used for the colour prints and holographic projections are available upon request.

**References**

1.  Matoba, O., Nomura, T., Perez-Cabre, E., Millan, M. S. & Javidi, B. Optical Techniques for Information Security. *Proc. IEEE* **97,** 1128–1148 (2009).

2.  Javidi, B. *et al.* Roadmap on optical security. *J. Opt.* **18,** 1–39 (2016).

3.  Knop, K. Color pictures using the zero diffraction order of phase grating structures. *Opt. Commun.* **18,** 298–303 (1976).

4.  Kumar, K. *et al.* Printing colour at the optical diffraction limit. *Nat. Nanotechnol.* **7,** 557–561 (2012).

5.  Sun, S. *et al.* All-dielectric full-color printing with TiO2 metasurfaces. *ACS Nano* **11,** 4445–4452 (2017).

6.  Lohmann, A. W. & Paris, D. P. Binary Fraunhofer holograms, generated by computer. *Appl. Opt.* **6,** 1739–1748 (1967).

7.  Goh, X. M. *et al.* Three-dimensional plasmonic stereoscopic prints in full colour. *Nat. Commun.* **5,** 1–8 (2015).

8.  Heydari, E., Sperling, J. R., Neale, S. L. & Clark, A. W. Plasmonic Color Filters as Dual-State Nanopixels for High-Density Microimage Encoding. *Adv. Funct. Mater.* **27,** 1–6 (2017).

9.  Huang, Y. W. *et al.* Aluminum plasmonic multicolor meta-hologram. *Nano Lett.* **15,** 3122–3127 (2015).

10. Zhao, W. *et al.* Full-color hologram using spatial multiplexing of dielectric metasurface. *Opt. Lett.* **41,** 147 (2016).

11. Wang, B. *et al.* Visible-frequency dielectric metasurfaces for multiwavelength achromatic and highly dispersive holograms. *Nano Lett.* **16,** 5235–5240 (2016).

12. Wan, W., Gao, J. & Yang, X. Full-color plasmonic metasurface holograms. *ACS Nano* **10,**

10671–10680 (2016).

13. Li, X. *et al.* Multicolor 3D meta-holography by broadband plasmonic modulation. *Sci. Adv.* **2,** e1601102–e1601102 (2016).

14. Barton, I. M., Blair, P. & Taghizadeh, M. R. Dual-wavelength operation diffractive phase elements for pattern formation. *Opt. Express* **1,** 54–59 (1997).

15. Bengtsson, J. Kinoforms designed to produce different fan-out patterns for two wavelengths. *Appl. Opt.* **37,** 2011–20 (1998).

16. Levy, U., Marom, E. & Mendlovic, D. Simultaneous multicolor image formation with a single diffractive optical element. *Opt. Lett.* **26,** 1149–51 (2001).

17. Lee, S.-H. & Grier, D. G. Robustness of holographic optical traps against phase scaling errors. *Opt. Express* **13,** 7458–7465 (2005).

18. Mohammad, N., Meem, M., Wan, X. & Menon, R. Full-color, large area, transmissive holograms enabled by multi-level diffractive optics. *Sci. Rep.* **7,** 1–6 (2017).

19. Nawrot, M., Zinkiewicz, Ł., Włodarczyk, B. & Wasylczyk, P. Transmission phase gratings fabricated with direct laser writing as color filters in the visible. *Opt. Express* **21,** 31919 (2013).

20. Gissibl, T., Wagner, S., Sykora, J., Schmid, M. & Giessen, H. Refractive index measurements of photo-resists for three-dimensional direct laser writing. *Opt. Mater. Express* **7,** 2293 (2017).

21. Makowski, M. *et al.* Simple holographic projection in color. *Opt. Express* **20,** 25130 (2012).

22. Gerchberg, R. W. & Saxton, W. O. A practical algorithm for the determination of phase from image and diffraction plane pictures. *Optik (Stuttg).* **35,** 237–246 (1972).

**Supplementary Information for "Holographic Colour Prints: Enhanced Optical Security by Combined Phase and Amplitude Control"**

Materials

All solvents were purchased from Sigma-Aldrich and used as-is. Photoresist (IP-dip, Nanoscribe GmbH) and glass substrates (fused silica, 25 mm squares with a thickness of 0.7 mm) were purchased from Nanoscribe GmbH.

Direct laser writing

Direct laser writing was performed in a Photonic Professional GT system (Nanoscribe GmbH). A 780 nm femtosecond pulsed IR laser with 90 fs pulse duration and 80 MHz repetition rate (Toptica FemtoFiber Pro) was focused into a liquid drop of IP-dip photoresist by an immersion objective (Zeiss Plan Apo 63×, NA 1.4) to induce two-photon absorption and polymerisation. The lateral position of the laser spot was controlled by using galvanometric mirrors to deflect the beam within the field of view of the objective lens, whereas the axial position of the spot was controlled by using a piezoelectric translation stage to shift the photoresist/substrate interface relative to the focal plane of the objective lens. This created pixels and prints as cross-linked polymer structures on the glass substrate.

The laser power incident on the entrance aperture of the objective lens was controlled by an acousto-optic modulator (AA Opto-Electronic). For line exposures (blocks), the scan speed was 8000 μm/s and the laser power 21.0 mW for the first raster scan and 16.8 mW for the second raster scan. The hatch linewidth was 250 nm and slice thickness 0.70 μm. For point exposures (pillars), the exposure time was varied between 0.02 and 0.04 ms and the laser power between 33.3 and 46.4 mW, and the slice thickness ranged from 0.69 to 1.01 μm. The slice thickness was adjusted to match the (dose-dependent) axial elongation of the point spread function of the laser spot while maintaining a vertical overlap of approximately 30% (300 to 430 nm depending on the size of the laser spot in the vertical direction).

To wash away the excess unexposed liquid photoresist, development was carried out by immersion of the sample in polyethylene glycol methyl ether acetate (PGMEA) for 5 minutes and then isopropyl alcohol (IPA) for 3 minutes, followed by transfer into nonafluorobutyl methyl ether (NFBME) as a low surface tension solvent for the final drying step. Due to the large difference in density between the two solvents, residual IPA carried over from the previous step would float on the surface of NFBME and had to be siphoned off before removing the sample. This was necessary to minimise recontamination of the sample with IPA when it was withdrawn through the surface, as the IPA would otherwise dry on the sample and cause the pillars to collapse due to its relatively high surface tension.

Thickness measurements

A series of blocks fabricated with different thicknesses in the range 0.6–1.8 μm was scanned with a stylus profilometer (KLA Tencor) at a lateral speed of 10 μm/s and a force of 0.10 mg to calibrate the thickness of the phase plates. Based on this study we estimated the limit of placement accuracy of the laser spot to be 100 nm in the axial direction. As such, we used thickness step sizes of no smaller than 100 nm for the phase plates, corresponding to a phase accuracy of 0.25π, 0.21π, and 0.17π, or approximately π/4, π/5, and π/6 (8, 10, and 12 phase levels) for design wavelengths of 449, 527, and 638 nm respectively. Because we applied a strict lower limit of 100 nm on the thickness step size, the actual number of phase levels used in the final prints became 7, 9, and 11 respectively.

Optical characterisation

Optical micrographs and spectra were acquired in an optical microscope (Nikon Eclipse LV100ND) equipped with a microspectrophotometer (Craic 508 PV). Samples were backlit by halogen lamp illumination and measured in transmission through a 5×/0.15 NA objective lens.

Spectra were analysed by comparing their average transmittance within three narrow wavelength bands centred around the red, green, and blue laser wavelengths, $\bar{T}_R$, $\bar{T}_G$, and $\bar{T}_B$. For each spectrum, these values were used to calculate figures of merit $\chi_R$, $\chi_G$, and $\chi_B$ that assess the suitability of the corresponding pillar array for use as a red, green, or blue colour filter. For example, the figure of merit for a red colour filter $\chi_R$ is the sum of the difference between red and green transmittances and the difference between red and blue transmittances, i.e. $\chi_R = (\bar{T}_R - \bar{T}_G) + (\bar{T}_R - \bar{T}_B) = 2\bar{T}_R - \bar{T}_G - \bar{T}_B$. A matrix representation was used in the code to allow for vectorisation of the actual computations.

## Photography of holographic projections

Holograms were projected using 638 nm red, 527 nm green, and 449 nm blue diode lasers (ThorLabs) onto a white wall at a distance of ~1 m away from the sample. The holographic projections, measuring ~10 cm in size, were imaged using a DSLR camera in a dark room.

## Scanning electron microscopy

Scanning electron micrographs were acquired in a field emission scanning electron microscope (JEOL JSM-7600F) at an accelerating voltage of 5.0 kV and a working distance of 6.7 mm.

## Simulations

The holographic projections in Fig. S2 were simulated based on the output of our design algorithm (a phase map and an amplitude map) given an input of the same three grayscale images from Fig. 1 but with different pixel arrangements. The simulated projections were calculated from the phase and amplitude maps in MATLAB as follows. First, the element-wise product of the phase map (in phasor form) and the amplitude map is computed, which approximates the initial electric field distribution of light just after being transmitted through the print. Then a Fourier transform is taken, approximating propagation of light into the far field (in the Fraunhofer limit). Lastly, the square modulus is taken to convert the electric field strength into an intensity image. The image is shown in logarithmic scale to approximate the human visual response.

## Colour palette

A representative selection of the colours attainable with the pillar array geometry using various combinations of the pillar dimensions (height and diameter) is shown in Fig. S1. Pillars on blocks (Fig. S1b) generally give darker colours that are spectrally shifted with respect to the colours of pillars patterned directly on the glass substrate (Fig. S1a). Note also that there is little to no colour variation within each patch of Fig. S1b despite the thickness variation of the underlying blocks.

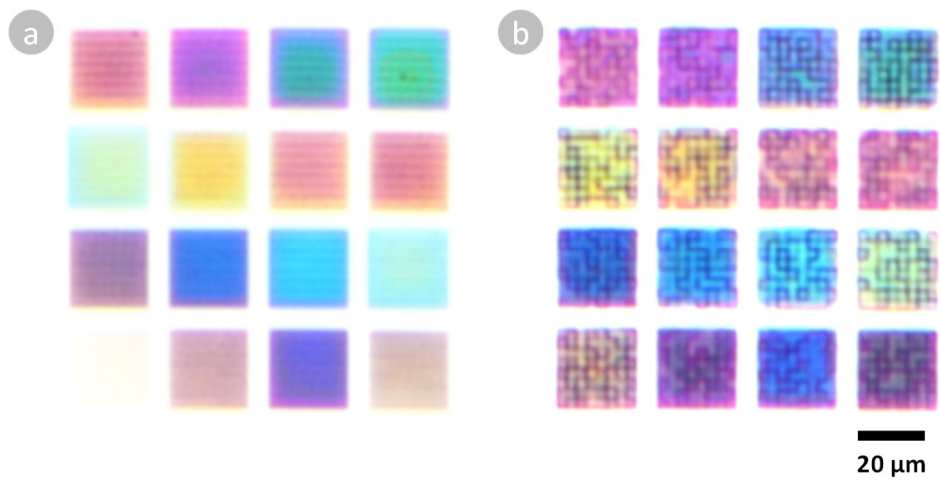

*Figure S1.* Colour palettes derived from pillar array colour filters by varying the pillar dimensions of height and diameter: (a) without blocks underneath (i.e. with a block thickness of 0 μm), and (b) with blocks of random thicknesses in the range 0.6–1.8 μm under the pillars.

Area allocation in space-division multiplexing of holograms

In a multiplexed hologram, the transmission efficiency for a given channel is the product of the area fraction occupied by the channel and the weighted average of the transmittance of the colour filters on that channel, with an upper bound of 33% for the case of equal area fractions in an RGB hologram. If unequal area allocation arises from a predominance of one or two colours in the colour image to be printed, this can be compensated by adjusting the colour balance of the image before colour matching. Alternatively, it can also be desirable to deliberately encourage an unequal area allocation when the wavelength selectivity of the filters on one channel is significantly worse than those on others. In this way, the number of the total hologram pixels allocated to each channel can be adjusted to balance out the transmission characteristics of the filters. For example, if the desired green transmission of the green filters does not sufficiently exceed the unwanted green transmission of the red and blue filters, more green pixels can be allocated to increase the signal-to-noise ratio on the green channel.

Indeed, we found that the quality of holographic projections was optimised by applying a slight green tint to the colour balance of the source image for the *Perfume* print in Fig. 4 to give a desirable pixel allocation of 36% green channel, 29% red channel, 27% blue channel, and 8% yellow pixels (not assigned to any channel). This green tint is not obvious in the final printed image, but is only used in the design stage to promote a desired colour matching outcome.

Pixel arrangements for space-division multiplexing of holograms

Holographic colour prints lie on a continuum between colour images, in which the arrangement of pixels is rigidly defined, and multiplexed holograms, for which the arrangement of pixels is seemingly arbitrary. However, even if the requirement to form a colour image is removed, there are still restrictions on the types of pixel arrangements that can be used for hologram multiplexing. Because our holograms are Fraunhofer holograms that operate in the Fourier domain, the Fourier transform of the (real space) pixel arrangement enters into the determination of the final holographic projections – specifically, *the final holographic projection is the spatial convolution of the designed holographic projection with the Fourier transform of the pixel arrangement*. The implications of this mathematical relationship on the design of multiplexed holograms are elaborated on in the following.

Adopting an idealised matrix representation of the pixel arrangement, the presence or absence of a hologram pixel at each location in space is denoted respectively by an amplitude of one or zero in the corresponding position of a 2D matrix (a binary mask). Then a matrix of ones corresponds to a hologram that completely fills the illuminated area and diverts the entire incident beam to project a desired image, while a matrix of zeroes corresponds to an illuminated area unoccupied by hologram pixels such that the incident beam passes straight through and remains as a spot.

In our space-division wavelength-multiplexing scheme, the pixels of each hologram only occupy part of the total area, which gives a "patchy" pixel arrangement on each wavelength channel. When pixels are removed from a complete, unmultiplexed hologram to create a patchy pixel arrangement (introducing zeroes into a matrix of ones), the undiffracted central (zero-order) bright spot increases in intensity at the expense of the projected image. While the projected image might then simply be expected to fade away gradually as pixels are removed, it can in fact become blurred or repeated. This is because the holographic projection is not only affected by the number of pixels remaining, but is also highly sensitive to the locations of the remaining pixels (i.e. the pixel arrangement).

To understand this, consider that the Fourier transform of a constant amplitude profile (the pixel arrangement of a complete, unmultiplexed hologram) is a Dirac delta function and returns an identical projection after convolution. However, the Fourier transform of the pixel arrangement of a patchy hologram is a combination of a Dirac delta and some noise terms which draw power away from the Dirac delta. Convolution with such a "noisy delta" function has the effect of creating unwanted copies of the holographic projection ("ghost images") that weaken and distract from the desired central projection.

These disturbances to the holographic projection can become especially pronounced when the Fourier transform has localised regions of high intensity noise that concentrate the ghost images and make them more apparent – a particularly serious issue when imperfect wavelength selectivity causes them to appear as crosstalk on multiplexed channels. In general, any form of ordering or periodicity in a pixel arrangement will be manifested as clustering or peaks in its Fourier transform and thereby accentuate the crosstalk noise in multiplexed holographic projections, as shown in the simulated far field projections in Fig. S2. Compared with a random pixel arrangement which produces three projections with little crosstalk (Fig. S2a), a "blocky" pixel arrangement band-limits the Fourier power spectrum, which concentrates ghost images around the central projection and gives it a blurry appearance (Fig. S2b). Meanwhile, a periodic pixel arrangement creates regular peaks in the power spectrum, which causes the tiling of ghost images in the Fourier plane (Fig. S2c). The crosstalk, which is barely noticeable in Fig. S2a, becomes much more apparent in Fig. S2b,c as the overlapping of ghost images enhances their visibility not only in their own channels but also on other channels.

Based on the above analysis, the optimal pixel arrangements for multiplexing are those which can spread out the ghost images by diffusing the noise power uniformly across the entire frequency domain to create a flat power spectrum, or equivalently, by generating a white noise signal in real space. We found that a convenient way to achieve a white noise spectrum was to use a dithering algorithm (e.g. Floyd-Steinberg error diffusion) to perform the necessary colour matching between the colour image to be printed and the colour palette available. Apart from increasing the perceived colour accuracy beyond the results of simple best-fit algorithms (which improves the quality of the colour print), dithering also scrambles the colour pixels and minimises occurrences of large blocks of colour pixels, which helps in randomising the pixel arrangement on each colour channel and generating a flatter power spectrum more similar to that of white noise (which improves the fidelity of holographic projections).

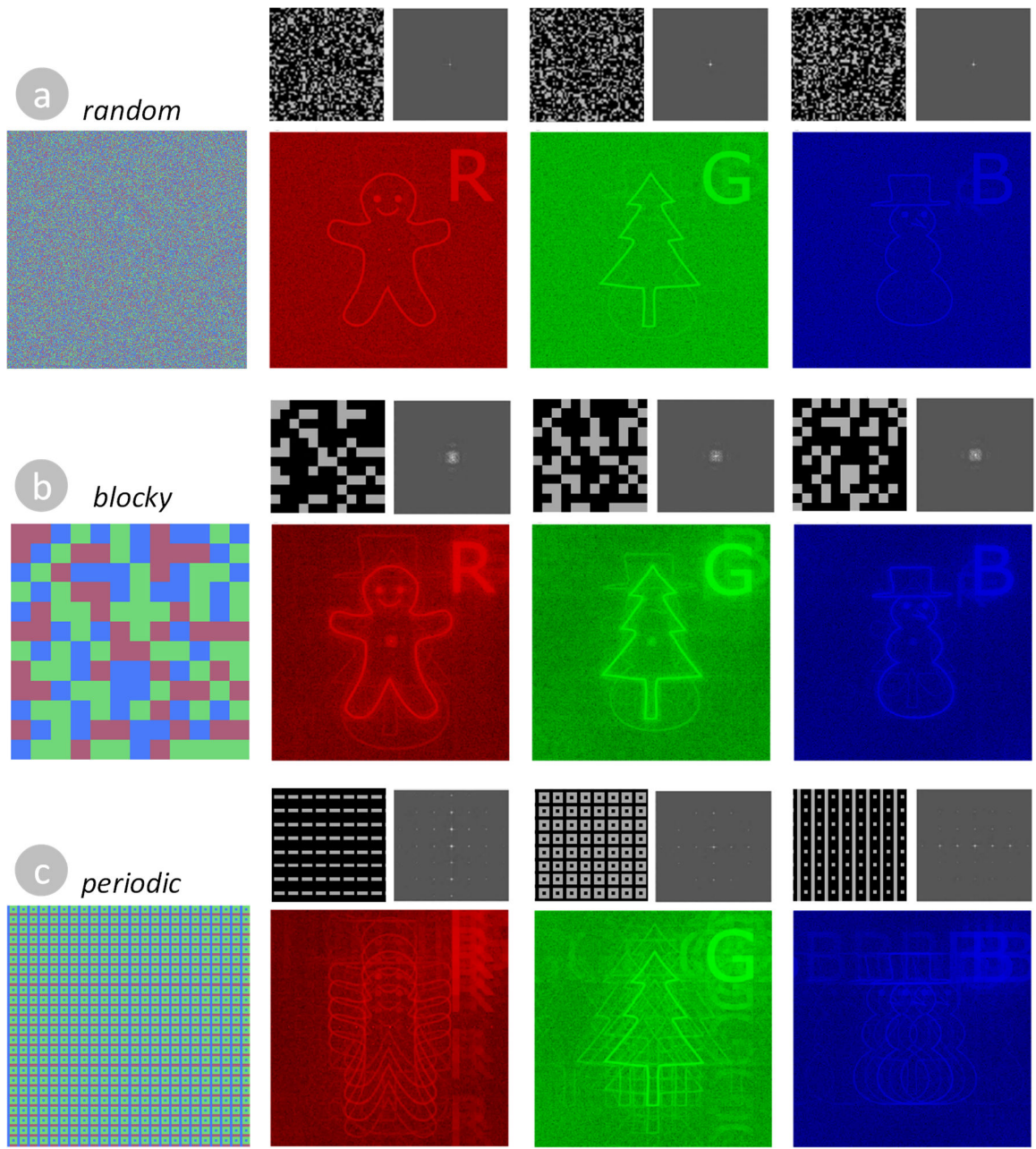

*Figure S2. Simulating the effect of pixel arrangement on hologram multiplexing.* A comparison of 480×480 red, green, and blue (RGB) pixel arrangements for hologram multiplexing and simulated far field holographic projections, based on the RGB laser wavelengths and transmission spectra in Fig. 2. Above each projection is the pixel arrangement for its colour channel, rescaled to show its essential features (left), and the Fourier transform of the pixel arrangement (right). (a) A random pixel arrangement in which individual RGB pixels are interspersed to give a featureless appearance. (b) A "blocky" random pixel arrangement in which randomness is only applied down to a scale of 40-pixel blocks. (c) A periodic pixel arrangement in which a 4×4 super-pixel is tiled to fill the space. The random

*arrangement (a) is the most suitable for hologram multiplexing as it accurately reproduces the source images with minimal crosstalk. With the blocky arrangement (b), unwanted "ghost images" are concentrated around the central holographic projection, forming a diffuse glow that highlights the crosstalk in the background. In the periodic arrangement (c), the ghost images are repeated across the Fourier plane at locations determined by the positions of peaks in the Fourier transform of the pixel arrangement.*